\documentstyle[preprint,aps]{revtex}
\begin{document}
\preprint{}
\draft
\title{Two-dimensional superfluidity and localization in\\
the hard-core boson model: A quantum Monte Carlo study}
\author{T. Onogi and Y. Murayama}
\address{Advanced Research Laboratory, Hitachi, Ltd.,
Hatoyama, Saitama 350-03, Japan}
\maketitle
\begin{abstract}
Quantum Monte Carlo simulations are used to investigate
the two-dimensional superfluid properties of the hard-core boson model,
which show a strong dependence on particle density and disorder.
We obtain further evidence that a half-filled clean system becomes
superfluid via a finite temperature Kosterlitz-Thouless transition.
The relationship between low temperature superfluid density and
particle density is symmetric and appears parabolic
about the half filling point.  Disorder appears to break
the superfluid phase up into two distinct localized states,
depending on the particle density.
We find that these results strongly correlate with the results
of several experiments on high-$T_c$ superconductors.
\end{abstract}
\hspace{0.5cm}
\pacs{PACS numbers: 74.20.Mn, 05.70.Jk, 72.15.Rn, 74.76.Bz\\
     \\
Submitted to Phys. Rev. Lett., June 3, 1993\\
Transferred/Resubmitted to Phys. Rev. B, December 21, 1993\\
Accepted for publication in Phys. Rev. B, December 30, 1993}
\newpage
\narrowtext

Quantum systems of interacting bosons in clean or disordered media
have been used to study a variety of interesting problems,
including the superfluidity of $^4$He,
superconductor-insulator transitions in thin films,
and vortex dynamics in type-II superconductors.
Recently,
the relation between superconducting transition temperature ($T_c$)
and carrier density ($\delta$) in various copper-oxide
superconductors was found to obey
a universal bell-shape curve \cite{uemura89,schneider92,zhang93}.
This finding also shows that the superfluid-like
condensation of  charged bosons (local Cooper-pairs)
is crucial for high-$T_c$ superconductivity,
based on some common features of copper oxides:
the extremely short coherence length implying
point-like Cooper pairs, and the layered structure which confines
the carriers mainly to two-dimensional (2D) CuO$_2$ layers.
{}From the analysis of $\mu$SR and transport measurements,
Schneider and Keller \cite{schneider92} argued that
such condensation of extreme type-II superconductors belongs to
the classical XY universality class in three dimensions,
and
Zhang and Sato \cite{zhang93} surmised that the bosons
involved are actually bipolarons.
In this model, Cooper-pair bosons whose binding energy is
much larger than condensation energy are assumed to
exist below $T_c$(onset), and superconductivity appears
via the phase coherence of pre-existing bosons.

Here, within the boson framework, we propose
a quantum XY universality class in two dimensions.
We examine a 2D lattice boson Hubbard model \cite{fisher89}
specified by  hard-core repulsive interactions and random potentials.
The phase diagram  depends on the temperature,
particle density, and also on the degree of disorder.
First and foremost, when the particle density is varied,
the phase diagram should be perfectly
symmetric about the half filling level,
because of the particle-hole symmetry.
Second, the hard-core bose gas has mathematical
analogy to the quantum spin problem \cite{matsubara56}.
{}From previous studies of spin-1/2 XY model
\cite{loh85,marcu87,onogi87,makivic92},
one may expect that, for a clean boson system,
the Kosterlitz-Thouless (KT) transition \cite{kosterlitz73}
takes place and 2D superfluidity  arises
below a finite critical temperature $T_{\rm KT}$.
Also, regarding different bosonic systems, there are early numerical
works suggesting a KT transition in a Coulomb gas model \cite{alder89}
and in a soft-core model \cite{krauth91a}.
Third, at sufficiently strong disorder,
dense bose gas is expected to exhibit a localized phase called
the Bose glass \cite{fisher89,runge92}.
Through a quantum Monte Carlo (MC) simulation,
this paper provides further support for the superfluid KT transition
of the hard-core boson model in the half filling case, and
sheds light on density-modulation and localization effects
in 2D superfluidity.
Our simulation results strongly correlate with
recent experimental data in copper-oxide superconductors, including
the $T_c$ vs. $\delta$ relationship and the nature of superconducting
phase transition.

The Hamiltonian of hard-core bosons is expressed as
\begin{equation}
 H_{\mbox{\scriptsize boson}}
     =  -{t \over 2}\sum_{\langle i,j \rangle}
       ({a_i}^\dagger{a_j} + {a_j}^\dagger{a_i})
                + \sum_{i=1}^{N} v_i n_i + H_{\mbox{\scriptsize HC}}
\label{hamiltonian}
\end{equation}
for a square lattice of size $L$$\times$$L$.
Here the first sum is the kinetic energy of bosons,
where the hopping constant $t=\hbar^2/2m^*_{\rm b} a^2$ with
effective mass $m^*_{\rm b}$ and lattice spacing $a$.
The second part represents the potential energy
from onsite disorders with a uniform distribution
$v \in [-\Delta, \Delta]$.
The hard-core interaction $H_{\mbox{\scriptsize HC}}$
inhibits the double occupancy of bosons at each site.
This interacting bose gas can be
transformed into an equivalent spin-1/2 XY system
in a randomly varying magnetic field, such that
$H_{\mbox{\scriptsize boson}} \rightarrow H_{\mbox{\scriptsize spin}}
 =-t\sum_{\langle i,j \rangle}
( S^x_iS^x_j+ S^y_iS^y_j)+ \sum_{i} v_i S^z_i$,
by setting $a^{\dagger}=S^x+i S^y $,
$a=S^x-i S^y$, and $n=S^z+1/2$ with
the spin-1/2 operator ${\bf S}=(S^x,S^y,S^z)$ \cite{matsubara56}.

To simulate the equilibrium state of such a qunatum
system efficiently,
we use the path-integral approach
based on the Suzuki-Trotter transformation \cite{marcu87,onogi87}.
The 2D bosonic system (under periodic boundary conditions)
is transformed to a 3D (torus with space and imaginary-time dimensions)
classical system of boson world lines.
Possible paths of the world lines are
topologically characterized by
particle number and winding number, where
the winding number ($W$) is defined by counting how many times
the world lines wind around the torus in space directions.
For the microcanonical ensemble of bosons
(fixed particle density $n_b$),
we carried out MC sampling of the world lines,
by utilizing an algorithm previously developed for the spin model.
( See Ref.~\cite{marcu87} for detail of the simulation method,
and Ref.~\cite{comment} for some technical remarks, respectively.)
The superfluid density $n_s$ corresponds to the free energy change
due to twisting the phase of order parameter along one lattice boundary,
and this quantity can be computed directly from
winding number fluctuations
\begin{equation}
n_s=(k_{\rm B}T/t) \langle W^2 \rangle
\label{ns}
\end{equation}
in the path-integral representation \cite{pollock87}.
For the MC sampling,
together with conventional local moves of world lines,
we also made global update of winding number $W$ \cite{alder89,krauth91a},
such that all the transitions to different values of $W$ are allowed.
In contrast, most previous spin simulations \cite{loh85,makivic92}
were limited to a fixed condition of $W=0$ while
the particle number was variable.
Since the present $W$-variable algorithm is ergodic regarding
winding number, we can evaluate $n_s$ exactly for a finite-size lattice.
This computational approach is complementary to earlier quantum MC studies
\cite{loh85,makivic92} and exact diagonalization study \cite{runge92}
on the similar model.

Figure~1 shows the numerical results of a clean half-filled
system ($\Delta=0$, $n_b=1/2$) with different size lattices ($L=4,6,8$).
The temperature dependence of 2D superfluid density
is shown in Fig.~1(a).
For $T<T_{\rm KT}= 0.40(5)\, t/k_{\rm B}$,
$n_s(T)$ grows remarkably and becomes
almost independent of the lattice size,
while, for $T>T_{\rm KT}$, it is suppressed with increasing size.
At low temperatures, $n_s$ reaches 0.27(1) for the $L=8$ lattice,
compared to 0.28 given by the exact diagonalization
of the much smaller ($L=4$) lattice \cite{runge92}.
Thus, even at the ground state, the superfluid fraction
is at most $n_s/n_b\approx 0.54$ in the thermodynamic limit,
and the normal-fluid component remains large.
This indicates a strong renormalization effect
due to quantum fluctuations,
while the thermal fluctuations cause
the vortex-antivortex pairs to unbind at high temperatures.
The value of $T_{\rm KT}$ is quite consistent with
the KT universal jump condition, which satisfies
$n_s(T_{\rm KT}) = 2k_{\rm B}T_{\rm KT}/\pi t$ denoted
as a dotted line in Fig.~1(a).
Moreover, to get a clearer sign of the KT transition,
we calculated the temperature derivative of $n_s$
using the fluctuation formula
\begin{equation}
\partial(\beta n_s)/\partial\beta
= \langle W^2 \rangle\langle E \rangle - \langle W^2 E \rangle
\label{diffns}
\end{equation}
where $E$ is the total energy and $\beta=1/k_{\rm B}T$.
As seen in Fig.~1(c),
the peak value of $\partial(\beta n_s)/\partial\beta$
grows rapidly near $T=T_{\rm KT}$, as the lattice size increases.
In sufficiently large lattices, Eq.~(3) reduces to the energy difference
$\Delta E$ between periodic and antiperiodic boundary conditions.
It should be noted that, for the classical XY model,
an analogously divergent sign of $\Delta E$
was used as a decisive evidence
for the KT transition \cite{himbergen81}.
On the other hand, the peak value of specific heat
tends to saturate at $C_v=0.62(1)$  just above
$T_{\rm KT}$ with the increase in size, as shown in Fig.~1(c).
Although it is well known that  free bose gas in two dimensions
never condenses at finite temperatures,
these numerical observations definitely
support the view  that
the {\it hard-core} nature of interactions between bosons
leads to the superfluid phase
through the finite temperature KT transition,
while the quantum fluctuations are rather significant at low temperatures.

Figure~2 shows how the superfluid density depends on
the bose-particle density and the degree of disorder ($\Delta$),
at low temperature $T = 0.25\, t/k_{\rm B}$.
Note that the relationship between the superfluid density $n_s$
and the doping density $n_b$ is approximated well,
although not exactly, by the parabolic form $n_s\propto (n_b-n_0)(1-n_b-n_0)$
with some constant $n_0$.  Here $n_0 =$ 0.06, 0.13, 0.13, and 0.15
for various degree of disorder $\Delta=$ 0, 1.0, 1.25, and 1.75, respectively.
In comparison, in a one dimensional (1D) pure system,
the exact result $n_s=\sin(\pi n_b)/\pi$ is obtained \cite{lieb61}.
The symmetry about half filling  is rigorous
in both clean and disordered systems,
since the model involves the particle-hole symmetry
due to hard-core interactions.
In the presence of disorder the parabola seems to be preserved, and
in addition the onset of superfluidity can be found
at a common value of $n_0\approx 0.1$, despite large reduction
of maximum $n_s$ at the half filling.
Since the transition temperature $T_{\rm KT}$ is
generally proportional to
$n_s(T=0)$ in the 2D system, the above parabolic property
implies that the relation between $T_c$ and particle density
takes a universal form,
which is accompanied with a common offset density in the disordered case.

The disorder always localizes bosons to suppress the superfluidity.
Nevertheless the disorder-induced localization appears to
differ qualitatively between dense and dilute bose systems
in two dimensions.
Figure~3 shows  the dependence of $n_s$ on  disorder ($\Delta$) at
a fixed low temperature.
The superfluidity of the moderately dense (half-filled) system is
insensitive to weak disorder. At stronger disorder $\Delta\gtrsim2.5$,
the superfluidity disappears, but the size dependence is very weak
even when the disorder is rather strong.
This result is consitent with earlier studies \cite{runge92,krauth91b}
which support the existense of the Bose galss phase for dense systems.
In contrast, the dilute system at 1/8-filling ($n_b=1/8$) is
rather sensitive even to weak disorder.  Here
$n_s$ is rapidly suppressed as the lattice size is increased.
This size dependence implies that the correlation length of superfluidity
is smaller than the lattice size $L=8$.
The insets of the figure
show the spatial distribution of boson density in each case.
For the dilute region in the vicinity of $n_b=n_0$,
most bosons appear to
form a superfluid cluster, as depicted in the upper inset.
The size of the cluster is comparable
to the localization length $\xi_l$ of the Anderson transition.
For instance,
using $\xi_l=a/\log(\Delta/t)$ in the limit of strong disorder,
$\xi_l\approx 5a$ for $\Delta=1.25$.
This observation indicates that the 2D weak localization is dominant
at densities below $n_0$.
On the other hand, for a half-filled system with strong disorder,
such clusters assemble into a percolative network,
as shown in the lower inset.
This behavior is supposed to be associated with critical fluctuations
in the vicinity of the Bose-glass transition point.
Here, when the glass correlation length,
larger than the Anderson's localization-length,
exceeds the lattice size,
the superfluidity disappears.
By varying the particle density under the hard-core condition,
we have thus obtained a possible tuning from Anderson glass to Bose glass
in two dimensions.  It is interesting
that the similar tuning was previously observed
in 1D disordered soft-core boson system ($n_b$=0.625) by varying
the strength of on-site repulsion from zero (free) to strong (hard-core)
couplings \cite{scalettar91}.

We made a  semi-quantitative comparison of these
numerical results with several results related to high-$T_c$ superconductors.
Hereafter we assume bosons to have a linear size
equal to the coherence length on the CuO$_2$ plane,
$a = \xi_{ab}\simeq10\AA$.
Recently Matsuda {\it et al.} \cite{matsuda92}
carried out transport measurements
which provided unequivocal evidence  of the KT transition in
a one-unit-cell(12$\AA$)-thick YBa$_2$Cu$_3$O$_{7-x}$ (YBCO) film
with a maximum value of $T_{\rm KT} =30$K.
The three-dimensional interlayer coupling, thus,
is not a necessary condition  for
finite-$T_c$ superconductivity,
although the interlayer coupling plays an additional role in
elevating the $T_c$ \cite{matsuda93}.
By comparing the experimental and numerically-determined
values of $T_{\rm KT}$ in the clean half-filled system
($2m^*_b{\xi_{ab}}^2 k_{\rm B}T_{\rm KT}/\hbar^2\simeq0.4$),
we estimate the boson mass as
$m^*_b=2m^*_{\mbox{\scriptsize carrier}}\simeq5.9m_e$,
which is comparable to the value of
$m^*_{\mbox{\scriptsize carrier}}\simeq 2.5m_e$
given by Schlesinger {\it et al.} \cite{schlesinger90}
from infrared measurements of YBCO.

In high-$T_c$ superconductors,
chemical doping does not only result in carrier-density modulation,
but also introduces some degree of microscopic disorder
in the superconductive CuO$_2$ plane, although the latter effect has not
been well elucidated experimentally.
One remarkable feature commonly observed in experimental data
is that the relationship between the transition temperature $T_c$ and
the hole density $\delta$ (per unit cell)
follows a common bell-type curve: $T_c(\delta)$
rises above $\delta_{\rm offset} \simeq 0.05$, reaches a maximum at
$\delta_{T_c(max)}\simeq 0.18$, and falls down to zero
at $\delta_{\rm end} \simeq 0.3$.
In addition, the $T_c$-$\delta$ curve is
symmetric about $\delta=\delta_{T_c(max)}$, possibly implying
some hidden symmetry.
These experimental features strongly correlate with the results
presented in Fig.~2.
Furthermore we theoretically
obtained the universal parabolic curve,
similar to the ansatz proposed in Ref.~\cite{schneider92}.
By simply assuming that the boson particle density scales with hole density
as $n_b=(\delta/2)(\xi_{ab}/l_{ab})^2$ with a unit cell size
$l_{ab}\simeq 4\AA$,
we get $n_b\simeq$0.16, 0.56, and 0.94 for
$\delta_{\rm offset}$, $\delta_{T_c(max)}$, and  $\delta_{\rm end}$,
respectively. This is also consistent with the computational results.
Thus, the existence of parabolic symmetry and the consistency of
characteristic values of $m^*_b$ and $\delta$
strongly suggest that superconductivity within individual CuO$_2$ layers
is caused by 2D superfluidity of local bosons moving
in somewhat disordered media.

To summarize, our quantum Monte Carlo simulations have provided further
support that the 2D hard-core bose gas undergoes a KT transition
to become superfluid,
and that the low temperature superfluid density
shows a parabolic-like dependence on particle density.
In disordered systems, we have shown that the bosons localize
differently in the dense and dilute cases.
The computational results correlate with
the $T_c$ vs. $\delta$ relationship for high-$T_c$ superconductors
and may  partially illustrate the nature of the superconducting transition.

We thank S. Nakajima, T. Ohtsuki, T. Uda, Y. A. Ono, and M. Hirao for
their discussions. T.O. thanks M. Suzuki, S. Miyashita, and M. Marcu
for their comments from theoretical viewpoint, and Y. Matsuda for
showing his experimental results prior to publication.

\begin{figure}
\caption{ (a) two-dimensional superfluid density $n_s$,
(b) temperature derivative of $n_s$, and (c) specific heat $C_v$,
as a function of temperature $T$, for a clean half-filled system.
The symbols denote the cases for different lattice sizes
$L=4$ (circles), 6 (triangles), and 8 (squares).
The solid curves represent spline fits to the calculated data.
The dotted line in (a) corresponds to the universal jump condition
of the Kosterlitz-Thouless transition.}
\label{fig1}
\end{figure}

\begin{figure}
\caption{ Low-temperature superfluid density as a function
of boson particle density, for different degrees of disorder.
This simulation was performed at low temperature $T=0.25\,t/k_{\rm B}$
on an $8\times8$ lattice.
Each value denotes the thermal and sample averages
over five different disordered states.
The degrees of disorder are denoted by circles for the
clean system ($\Delta=0$), triangles ($\Delta=1.0$),
squares ($\Delta=1.25$), and crosses ($\Delta=1.75$).
The solid lines represent parabolic fits to the calculated data.
}
\label{fig2}
\end{figure}

\begin{figure}
\caption{  Localization in
dense ($n_b=1/2$) and dilute ($n_b=1/8$) bose systems.
The curves show the low temperature
superfluid density  as a function of disorder strength,
at a fixed temperature $T=0.25\,t/k_{\rm B}$.
The symbols correspond to lattice sizes $L=4$ (black circles),
8 (black squares) for the dense system,
and $L=4$ (circles), 6 (triangles),
8 (squares) for the dilute system.
The two insets represent the density distribution of
bosons on the $8 \times 8$ lattice. The upper and lower insets
correspond to the cases of ($n_b=1/8$, $\Delta=1.0$) and
($n_b=1/2$, $\Delta=2.5$), respectively.
}
\label{fig3}
\end{figure}


\begin{references}
\bibitem{comment}
   To breakup the partition function according to the Suzuki-Trotter
transformation, we used the checkerboard type of real-space
decomposition of $H_{\mbox{\scriptsize spin}}$ according to
Refs.~\cite{loh85} and \cite{marcu87};
$Z={\rm Tr}\,e^{-\beta H} \approx
{\rm Tr}\,(e^{-\Delta\tau H_1} e^{-\Delta\tau H_2})^{\beta/\Delta\tau}$
with $n=\beta/\Delta\tau$ time slices, where
the Hamiltonian is decomposed into two kinds of
cell(four-spin)-hamiltonians $H=H_1+H_2$ such that they do not overlap.
This formula is accurate up to $O(\Delta\tau^2)$.
We fixed $\Delta\tau=1/4$, and
performed (1-4)$\times 10^4$ lattice sweeps after 10$^3$ sweeps of
thermalization, typically for a single run.
The systematic error in the Trotter approximation $\Delta\tau$
is smaller than the statistical error of a few percent.
\end{references}
\end{document}